\documentclass[aps,prd,preprint,superscriptaddress,tightenlines,nofootinbib]{revtex4}

\usepackage{graphicx}
\usepackage{dcolumn}
\usepackage{bm}
\usepackage{epsfig}
\begin{document}

\preprint{CLEO CONF 03-07}   
\preprint{LP-122}      

\title{Evidence for Two-Body Hadronic Decays of the $\Upsilon$\ }
\thanks{Submitted to the 
XXI International Symposium on Lepton and Photon Interactions at
High Energies,
August 2003, Fermilab, Batavia, IL, USA}

\author{S.A.~Dytman}
\author{J.A.~Mueller}
\author{S.~Nam}
\author{V.~Savinov}
\affiliation{University of Pittsburgh, Pittsburgh, Pennsylvania 15260}
\author{G.~S.~Huang}
\author{D.~H.~Miller}
\author{V.~Pavlunin}
\author{B.~Sanghi}
\author{E.~I.~Shibata}
\author{I.~P.~J.~Shipsey}
\affiliation{Purdue University, West Lafayette, Indiana 47907}
% ...
\collaboration{CLEO Collaboration} 
\noaffiliation

\date{July 15, 2003}

\begin{abstract} 

We describe a search for hadronic
decays of the $\Upsilon$(1S), $\Upsilon$(2S), and  $\Upsilon$(3S)
resonances to the exclusive final states
$\rho\pi$, $K^*(892)\bar{K}$, $\rho a_2(1320)$, 
$\omega f_2(1270)$, 
$\phi f_2'(1525)$, 
$K^*(892)\bar{K}^*_2(1430)$,
$b_1(1235)\pi$,  $K_1(1270)\bar{K}$,
and $K_1(1400)\bar{K}$. 
Upper limits at 90\%~CL are
set for all these decays from all three resonances
below $33\times 10^{-6}$; in particular,
$\cal{B}$($\Upsilon $(1S)$\to\rho\pi)~<~4\times 10^{-6}$
is the smallest such upper limit.
For two modes, a branching fraction of zero can be ruled out 
with a statistical significance of more than $5\sigma$: 
$B(\Upsilon(1S) \to \phi f_2'(1525)) 
= (7^{+3}_{-2} \pm 1) \times 10^{-6}$ and
$B(\Upsilon(1S) \to K_1(1400) \bar K) 
= (14^{+4}_{-3} \pm 2) \times 10^{-6}$.
Production of $K_1(1270) \bar{K}$ in $\Upsilon$(1S) decay is suppressed relative
to that of $K_1(1400) \bar{K}$. 
These results add another piece
to the challenging ``$\rho-\pi$ puzzle'' of the charmonium system,
placing constraints on models of how quantum chromodynamics
should be applied to heavy quarkonia.
All results are preliminary.
\end{abstract}

\vspace*{2mm}
\maketitle

\newcommand{\oneS}{\mbox{$\Upsilon(1S)$}}   
\newcommand{\twoS}{\mbox{$\Upsilon(2S)$}}  
\newcommand{\threeS}{\mbox{$\Upsilon(3S)$}}  
\newcommand{\fourS}{\mbox{$\Upsilon(4S)$}}

  
\section{Introduction and Motivation}  
Little experimental information exists on exclusive decays of the  
$\Upsilon$ resonances below $B\bar{B}$ threshold. 
Upper limits have been published for the decays  
$\Upsilon(1S)\to\rho\pi$ and $\Upsilon(1S)\to h\bar h$,   
$h=\pi,\ K,\ p$,   
all of the order $10^{-4}$,   
and $\Upsilon(1S) \to \pi^+ \pi^- \pi^0$ ($1.8\times  
10^{-5}$)~\cite{pdg2002}.   
No exclusive final states for the $\twoS$ and $\threeS$ have been examined.  
The situation is different in charmonium, where numerous channels have been  
measured.  
This by itself poses a motivation to study hadronic $\Upsilon$ decays, given  
the  
similarity of these two strongly-bound heavy quark systems.  
Furthermore, a long-standing unsolved puzzle in charmonium regarding the  
ratio of branching fractions  
of the $2^3 S_1$ ($\psi'$) and $1^3 S_1$ ($J/\psi$)   
to final states consisting of a pseudoscalar and   
a vector warrants the corresponding measurement in bottomonium.   
  
The expectation that the dilepton and hadronic ratios of decay widths should  
be at least  
roughly equal follows from QED and~QCD. Both processes are  
thought to  
occur via annihilation of the constituent quark and antiquark, in one case to  
a photon  
and in the other to three gluons, and therefore are both proportional to the  
square of the  
quark-antiquark wave function overlap at the origin. Restating the  
QCD~expectation for  
hadronic decays in terms of ratios of branching fractions instead of decay  
widths, and neglecting 
the running of the strong coupling constant\footnote{
The strong coupling constant $\alpha_s$ enters with the third power. The 
relevant ratio between the $\psi(2S)$ and the $J/\psi$ is 
$\left(\alpha_s(m_{\psi(2S)}) / \alpha_s(m_{J/\psi}) \right)^3=
0.85$~\cite{guandli}.
},  
one obtains the following prediction regarding an arbitrary final state~$H$:  
  
\begin{equation}  
Q=\frac{{\cal B}(\psi(2S) \to H )}{{\cal B}(J/\psi \to H)}   
\approx
\frac{{\cal B}(\psi(2S) \to e^+ e^- )}{{\cal B}(J/\psi \to e^+ e^-)}   
\label{eqn:Q}  
\end{equation}  
  
\noindent  
Using the leptonic branching fractions  
${\cal B}(J/\psi \to e^+ e^-) = (5.93 \pm 0.1 ) \times 10^{-2}$  
and
${\cal B}(\psi(2S) \to e^+ e^-) = 
(7.3 \pm 0.4) \times 10^{-3}$~\cite{pdg2002},  
the expected value\footnote{  
Using earlier measurements and setting the ratio of coupling constant values  
to unity, the ratio of dilepton branching resulted in a ratio $Q\sim 14\%$.  
This  
is the reason why the puzzle posed by the failure of some channels was  
referred to as ``the $14\%$~puzzle''.  
}   
for the ratio is $(12.3 \pm 0.9)\%$.

A number of channels have been studied, most of which satisfy the 
prediction within experimental errors.   
The most significant deviation known so far comes from 
the following vector-pseudoscalar~($VP$) 
and vector-tensor~($VT$) final states~\cite{pdg2002}:   
$\rho\pi$ ($Q< 0.007$), $K^*(892) \bar K$ ($Q<0.001$), $\rho a_2(1320)$  
($Q<0.02$),  
$K^*(892) \bar{K_2^*}(1430)$ ($Q<0.02$), and   
$\omega f_2(1270)$ ($Q<0.03$~\cite{BESPsiprimedecays,pdg2002}).   
 
Many theoretical approaches  
have been made to resolve this puzzle. None is able to accommodate all the  
measurements  
reported so far. For some, the crucial question is whether the  
$J/\psi$ is enhanced or the $\psi'$ is suppressed.   
Therefore, in addition to adding experimental information to the scarce amount  
that is available at this point on $J^{PC}=1^{--}$ bottomonium decays,   
in particular in the $\rho\pi$ channel, it is an interesting question   
what the ratio analog to Equation~\ref{eqn:Q} should turn out to be for   
the~$\Upsilon$ system.

Depending on which theoretical model is used to explain the behavior  
measured in  
charmonium, the expectation for bottomonium varies. A common assumption is  
that the  
individual branching fractions in bottomonium should be at least one order of  
magnitude  
smaller than in charmonium. How much they are smaller should be of  
some utility in deciphering the puzzle.  
 
Two important features distinguish the situation in bottomonium from that in  
charmonium.  
First, in addition to the comparison of the 2S excitation with the ground  
state,   
the~3S resonance can be used due to the fact that it is below open flavor  
production threshold, in contrast to the situation in charmonium.  
Furthermore, the ratio predicted based on   
Equation~\ref{eqn:Q} is $48\%$ for $\twoS : \oneS$ and $72\%$ for  
$\threeS : \oneS$.  
  
CLEO recently accumulated several million bottomonium decays at each of the  
$\oneS$, $\twoS$,  
and $\threeS$ resonances. These datasets can probe decays of the  
$J^{PC}=1^{--}$~$b\bar b$ bound  
states at the $10^{-5}$ level. The decays pursued in this work are   
$\rho\pi$,   
$K^*(892) \bar K$,   
$\rho a_2(1320)$,   
$\omega f_2(1270)$,   
$\phi f_2'(1525)$,   
$K^*(892) K_2^*(1430)$,   
$b_1(1235)\pi$,   
$K_1(1270) \bar K$, and   
$K_1(1400) \bar K$.  
They sample $VP$, axialvector-pseudoscalar~($AP$), 
and $VT$~type final states with and without strangeness and  
constitute the most copious  two-body  
hadronic final states in $J/\psi$~decay (each with a branching ratio of  
$0.1 - 1\%$).  
Each proceeds via the strong interaction and conserves isospin.   
These decay modes provide  
a logical starting point for using the $\Upsilon$ system to untangle the  
many subleties of the $\rho\pi$ puzzle and associated anomalies.  
  
\section{Analysis overview}  

The analysis strategy is straightforward. Event selection criteria for the  
different modes  
are developed using signal Monte Carlo. We emphasize cleanliness over  
efficiency in order  
to suppress decays faking the desired final states.  
CLEO data taken at and just below the $\fourS$  
resonance, suitably scaled by luminosity and beam energy, is used as an  
indication of   
background properties and final contamination levels.   
Finally, projected backgrounds and  
event totals are normalized by efficiencies and the number of resonance decays  
produced, and branching fractions or upper limits computed.   
  
It is important to note that many different kinds of backgrounds contribute  
to the sample  
obtained at the $\fourS$ resonance. Not all of them scale with the same 
center-of-mass energy dependence (see discussion below). 
For that portion of backgrounds which are truly the same  
final state but produced non-resonantly,  
that is, proceed as $e^+ e^- \to \gamma^* \to H$ instead of  
$e^+ e^- \to \Upsilon \to H$,  
there is also the possibility of interference, which has been
neglected in this work.
  
\section{Event Selection}  

The CLEO~III detector is described in detail in~\cite{cleoiiidetector}.  
Its key features  
exploited in this analysis are a solid angle
coverage for charged and neutral particles of 93\%   
and two particle identification systems 
to separate kaons from pions, namely using energy loss in the drift chamber  
and a Ring Imaging Cherenkov detector~\cite{RICHNIM}.
The tracking system achieves 
a charged particle momentum resolution of 0.35\% (1\%) at
$p=$1~GeV/$c$ ($p=$5~GeV/$c$) and the calorimeter a photon 
energy resolution of 2.2\% (1.5\%) at $E_\gamma=$1~GeV ($E_\gamma=$5~GeV).
The combined $dE/dx$-RICH particle identification system attains
a kaon efficiency (fake rate) $>$90\% ($<5$\%) below 2.5~GeV/$c$ and falls
(rises) to $\sim$70\% ($\sim$25\%) near 5~GeV/$c$.

Standard requirements are used to identify charged particles from tracks in the 
drift chamber and photons from electromagnetic showers in the CsI calorimeter.  
The simplicity of the final states under study is exploited by  
imposing an energy conservation requirement on
$X_T=\Sigma_h(E_h)/(2E_{beam})$ 
of $0.98 \leq X_T  \leq 1.015$ when   
summing over the energies of the decay products. 
For all the target modes, the experimental resolution in this quantity is
smaller than~$1\%$.
For each of the final state resonances, a search window for the invariant  
mass of the   
decay products is established based on signal Monte Carlo studies.  
Electron and muon vetoes are imposed to suppress QED~backgrounds.  
  
\vspace{-0.5mm}
  
\section{Monte Carlo samples}  

Efficiencies are evaluated with Monte Carlo simulation of  
the process and detector response~\cite{GEANT}.  
The $\rho\pi$ and $K^* \bar K$ modes are generated with the polar   
angle~$\theta$  
distributed according to $(1+\cos^2 \theta)$. All other channels are thrown  
to be flat  
in $\cos \theta$ as they can be of any linear combination of 
$\sin^2\theta$ and  
($1+\cos^2\theta$). 
The final efficiencies are of the order of $5 - 10\%$,  
including  
all effects of selection criteria and all intermediate branching fractions   
(Table~\ref{tab:efficiencies}).  
The detection efficiency of some of  
the modes varies significantly with the beam  
energy and so is evaluated at the  
$\oneS$ and $\fourS$ separately. The $\twoS$ and $\threeS$ efficiencies,
also listed in Table~\ref{tab:efficiencies}, are obtained by interpolating  
linearly between the $\oneS$ and $\fourS$.  
 
Systematic uncertainties on the efficiency include uncertainties in the 
polar angle distribution for $VT$ and $AP$ modes (5\%) 
and modeling of tracks (1\% per charged track), $\pi^0$s (8\% per beam-energy-$\pi^0$
and 5\% per softer $\pi^0$), lepton veto (1\% per particle),  
kaon identification (5\% per identified kaon) and pion fake rate (3\%),  
and secondary vertex-finding (5\%).  
These contributions
and the uncertainty in the number of produced  
$\Upsilon$~decays ($5\%$) are summed in quadrature. The resulting 
total relative error is close to $10\%$ for all modes. 
 
Although $\tau$-pair production of the states in question 
contributes for $X_T < 0.98$, 
the background in the signal region is found to be small, 
as is that from $\mu$-pairs.

\vspace{-0.5mm}
  
\section{Determination of signal and background yield}  

\begin{figure}[thp]
\caption{Scaled total energy, $X_T$, distribution for candidate
channels from $\Upsilon$(1S), in which all selection criteria
$except$ the $X_T$ requirement have been applied. 
Solid points represent data; the red 
(dark) histogram is signal MC with arbitrary normalization; the 
green (light) histogram is scaled 4S data.\label{fig:xtot} }

\includegraphics*[width=1.3\textwidth]{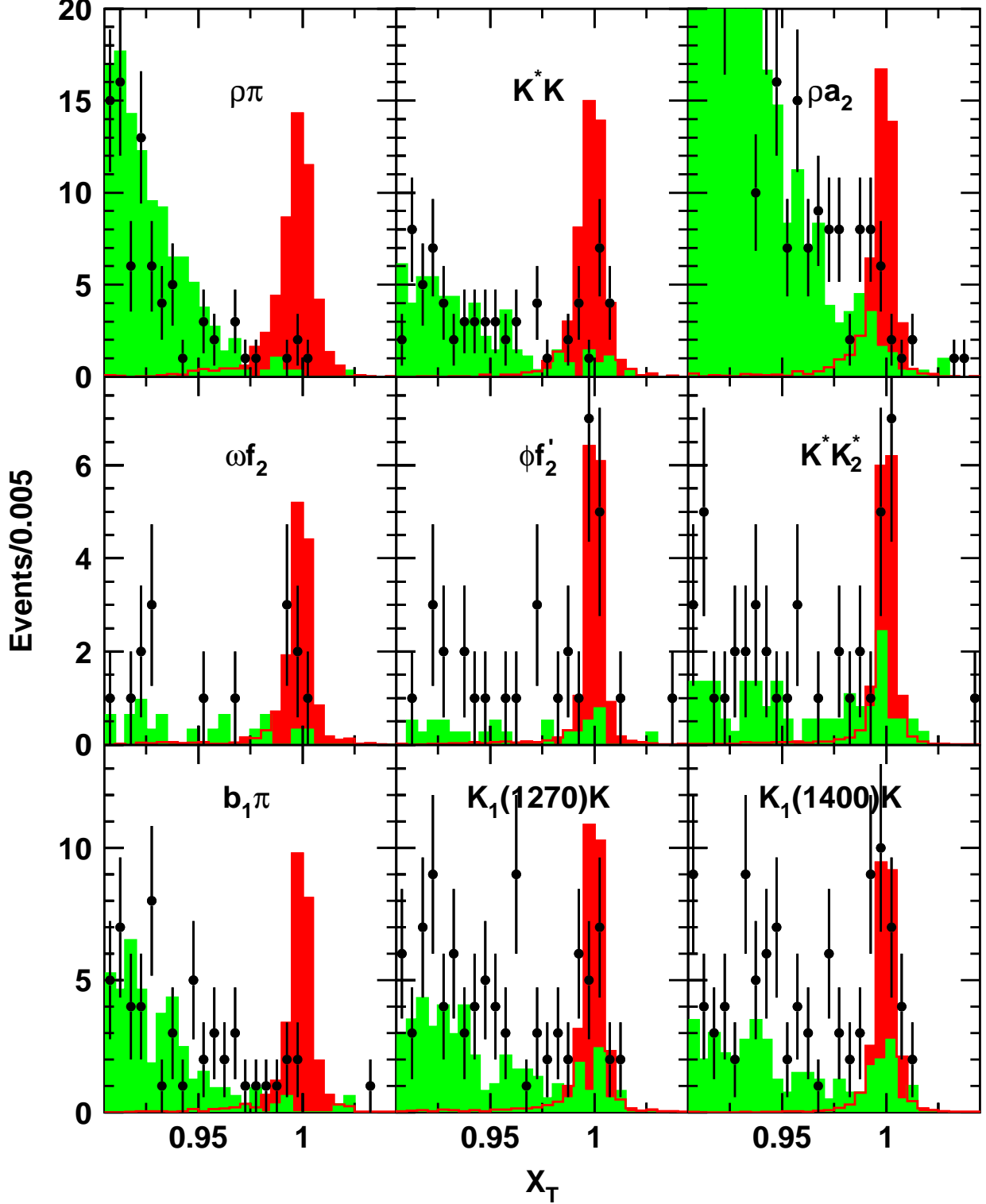}

\end{figure}

\begin{figure}[thp]
\caption{Mass distributions of intermediate states $M_X$, 
with $X$ given by the label in parentheses, for candidate
channels from $\Upsilon$(1S) decays, in which all selection criteria 
$except$ the one on
the plotted mass have been applied. Solid points represent data; the red 
(dark) histogram is signal MC with arbitrary normalization; the 
green (light) histogram is scaled 4S data. Plots on the right refer to the same
channel as each neighboring plot on the left.\label{fig:mx} }

\includegraphics*[width=1.3\textwidth]{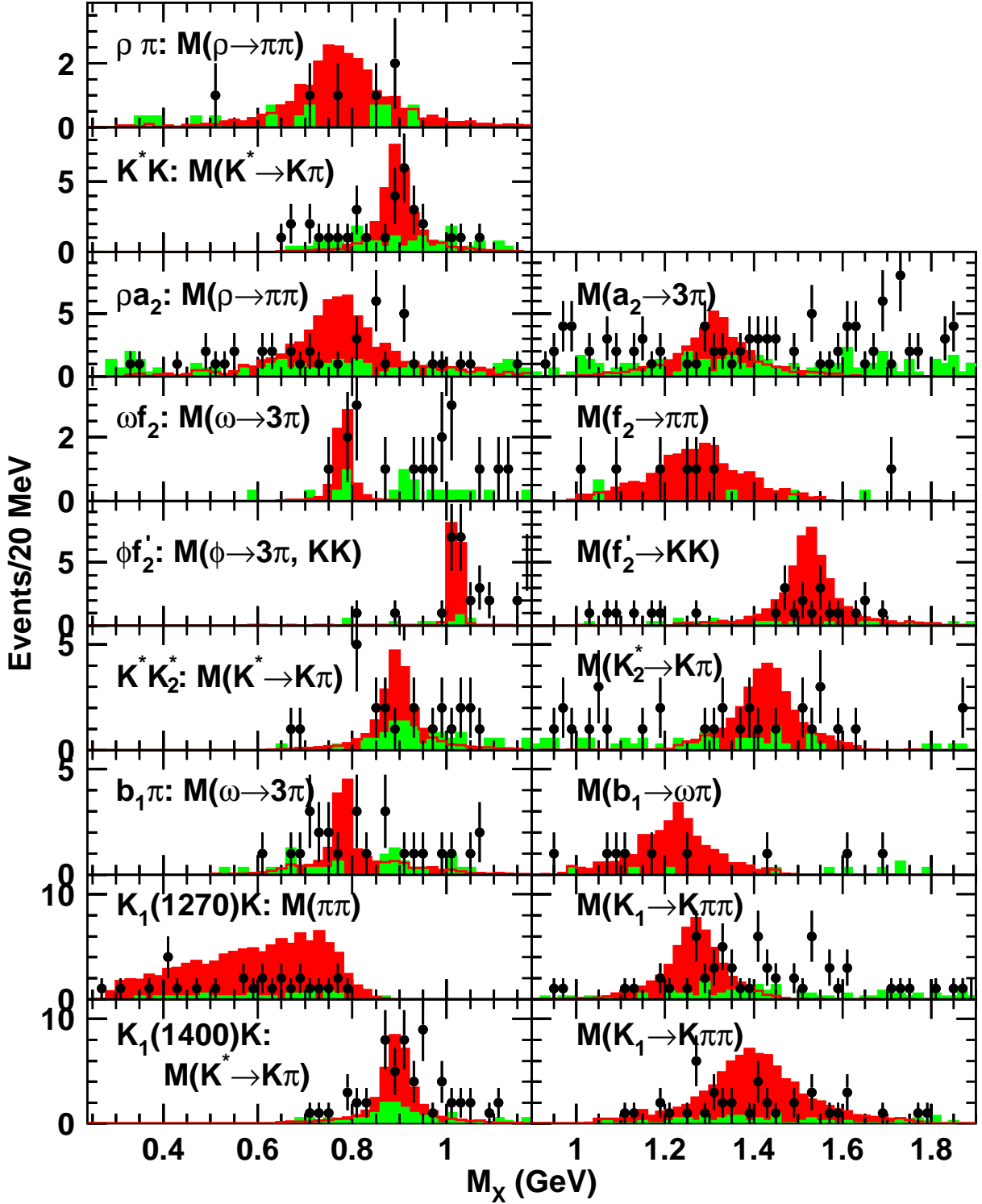}

\end{figure}

The data samples used consist of  
($21 \pm 1$), ($5.4\pm 0.2$), and ($5.0 \pm 0.3$) $\times 10^6$ 
$\oneS$, $\twoS$, and $\threeS$ decays, respectively.\footnote{ 
The number of resonance decays was computed using preliminary 
hadronic cross section line shape measurements with CLEO~III by scanning 
the center-of-mass energy around the resonance. 
} 
Events that satisfy the event  
selection criteria mentioned in the previous section are simply counted.  
To obtain maximal statistical power, 
both charged and neutral decay modes were  
considered and combined for the final result.   
The  
uncorrected event numbers thus obtained
are listed in Table~\ref{tab:signalyield}.  
Example distributions of the scaled total energy and invariant masses  
for the nine channels under study for \oneS~decays can be found in  
Figures~\ref{fig:xtot} and~\ref{fig:mx}.   
 
The background level was determined using a large amount of data taken on or near the  
$\fourS$ resonance, extrapolated down to the lower resonances. For the extrapolation,  
three issues must be taken into account:  
Scaling with luminosity, efficiency dependence on  
the center-of-mass energy, and cross section dependence on the center-of-mass energy.  
The efficiency dependence can be read off Table~\ref{tab:efficiencies}. 
The ratio of integrated luminosities for the datasets used is 
0.133, 0.093, and 0.135 for  
$\oneS:\fourS$, $\twoS:\fourS$, and $\threeS:\fourS$,   
respectively. The cross section extrapolation with beam energy 
poses the most uncertain contribution. Depending on the contributing
background process,  
it could vary from $1/s$ for QED processes to as much as $1/s^3$ ($1/s^4$ for  
$VP$~channels   
such as $\rho\pi$ or $K^* \bar K$).\cite{BRODLEP} 
We therefore quote ranges of scale  
factors that take this uncertainty into account, shown in Table~\ref{tab:scalefactors}.  
The scale factors vary between~$0.10$ and~$0.45$. 
Within a specific channel,  
the ratio of upper to lower scale factor limit ranges from a maximum of 
almost~$2$ for the $\oneS$ $VP$~modes 
down to the level of $1.1$ for $\threeS$ channels.  

Cross-feed between the investigated final states 
is accounted for as a separate source of background. 
The largest contribution is 
found to be $K_1(1400)\bar K$ events leaking into the 
$K_1(1270)\bar K$ sample. 
The $K_1(1400)\bar K$ signal above $M(K\pi\pi)=1.4~GeV$ (where there is no
$K_1(1270)$ signal) is evaluated with only the $\fourS$ background considered
and then, scaled in accordance to the Monte Carlo detection efficiency,
taken as a second source of background for the $K_1(1270)\bar K$.
We neglect cross-feed in the other direction because the predicted backgrounds
to $K_1(1270)\bar K$ saturate the observed rate.
Also, the $\rho\pi$ channel receives contributions from mis-identified $K^*(892)\bar K$ 
events in the $\rho^- \pi^+$ mode. Again, the $K^*(892)\bar K$ leakage is 
treated as a second source of background, properly scaled. 
Using the $\fourS$~event yield from Table~\ref{tab:signalyield} 
together with the maximal scale factors from Table~\ref{tab:scalefactors},  
and, where necessary, scaled cross-feed contributions, 
one arrives at estimates    
of background levels listed in Table~\ref{tab:backgroundlevel}.  
  
The confidence level that any given mean signal combined with background 
would exceed or equal the observed event count is computed from simulated trials 
in which a pseudo-random number generator is employed to throw Poisson 
distributions. Poisson fluctuations in both the observed resonance and 
4S samples are simulated
by allowing not only the background to vary around its mean 
from one trial to the next, but also
the mean background itself: 4S levels are fluctuated around the
observed number prior to application of the scale factor to obtain the
mean, and only then are fluctuations on the mean introduced. 
We reject thrown backgrounds which exceed the number of 
observed events. Trials are thrown in steps of~0.1 in signal mean until the 
desired confidence level is exceeded.  
This procedure predicts slightly larger intervals than the 
approach by Feldman and Cousins~\cite{FELDCOUS} when  
backgrounds are less than the observed number of events, and 
considerably wider ones for observations smaller than the 
mean expected background.  
  
Upper limits at $90\%$~CL on the number of signal events are listed in   
Table~\ref{tab:nrsigevtsupperlimits}, for which the $1/s$~scale factors
were used to minimize the estimated non-resonant background and
therefore to maximize any potential signal. These are converted into $90\%$~CL  
upper limits on the corresponding branching fractions shown in 
Table~\ref{tab:tableLast} using the 
number of produced resonance decays and efficiencies
(Table~\ref{tab:efficiencies});
we account for the systematic relative error of $10\%$ in this conversion
by increasing each upper limit by an additional $1.28\sigma\cong 13\%$.  

Two-sided intervals of $68\%$~confidence level are also
shown in Table~\ref{tab:tableLast} for channels with statistical
significance
exceeding one standard deviation.  
We define the statistical significance
to be the number of Gaussian standard deviations above which
lies the probability that the background alone fluctuated up to the
observed number of events.
For these intervals, the upper end of the 4S scale factor range
is employed to minimize the chance of undersubtracting background.
The systematic error shown includes the uncertainty on the product
of efficiency and number of $\Upsilon$'s produced mentioned above
(10\%) in quadrature with an additional 10\% to account for
uncertainties in $e^+e^-$ annihilation backgrounds as well as
$\Upsilon$ cross-feed estimates.
For two channels, $B(\oneS \to \phi f_2'(1525))$ and 
$B(\oneS \to K_1(1400) \bar K)$,
zero branching fraction can be ruled out at a statistical significance of 
more than $5\sigma$,
making these the first exclusive hadronic decay modes measured in 
the $\Upsilon$~system.
Several other channels with significance near $\sim 3\sigma$ show 
suggestive but statistically marginal evidence for branching fractions
at the few per million level.

In contrast to the $\psi(2S)$ and in similarity to the 
$J/\psi$~\cite{BESpsi'}, the
production of $K_1(1270) \bar{K}$ is suppressed relative
to that of $K_1(1400) \bar{K}$. 

\section{Conclusions}  
Using CLEO~III datasets of 21, 5.4, and 5.0 million $\Upsilon$(1S),  
$\Upsilon$(2S), and $\Upsilon$(3S) decays, respectively, we have searched for  
nine of the potentially most probable two-body-hadronic decays,  
which include $VP$, $AP$, and $VT$ channels: 
$\rho\pi$,   
$K^*(892) \bar K$,   
$\rho a_2(1320)$,   
$\omega f_2(1270)$,   
$\phi f_2'(1525)$,   
$K^*(892) K_2^*(1430)$,   
$b_1(1235)\pi$,   
$K_1(1270) \bar K$, and   
$K_1(1400) \bar K$.  
The upper limit at 90\% confidence level for  
$\Upsilon$(1S)$\to\rho\pi$ is lowered by more than an order of  
magnitude to $4\times 10^{-6}$,   
and 90\% CL upper limits for the other eight modes, 
measured for the first time, range from  
$7 - 33\times 10^{-6}$. 
Two channels have been observed at convincing statistical significance:
$B(\Upsilon(1S) \to \phi f_2'(1525))
= (7^{+3}_{-2} \pm 1) \times 10^{-6}$ and
$B(\Upsilon(1S) \to K_1(1400) \bar K)
= (14^{+4}_{-3} \pm 2) \times 10^{-6}$.
The branching fractions
from the $\Upsilon$(1S) are smaller than the comparable values 
on the $J/\psi$ by factors of several hundred (for $K_1(1400) \bar{K}$) to at least 
several thousand (for $\rho\pi$); the $\Upsilon(1S)\to \rho\pi$ 
branching fraction is measured to
be suppressed by at least six powers of $M_{J/\psi}/M_\Upsilon$
relative to $J/\psi\to \rho\pi$. The above results are preliminary.

\begin{acknowledgments}
We would like to thank Stan Brodsky, Eric Braaten, Henry Tye, and Nicholas
Jones for motivating and illuminating communications.
We gratefully acknowledge the effort of the CESR staff
in providing us with
excellent luminosity and running conditions.
This work was supported by
the National Science Foundation,
the U.S. Department of Energy,
the Research Corporation,
and the
Texas Advanced Research Program.

\end{acknowledgments}

  
\clearpage  
 
\clearpage  

\begin{table}[!htp]
\setlength{\tabcolsep}{0.4pc}
\catcode`?=\active \def?{\kern\digitwidth}
\caption{Efficiency in percent
for the isospin- and charge-conjugate-inclusive parent
decay channels listed, summed
over the various sub-modes used in this analysis, including
all the effects of selection criteria and all
intermediate branching fractions.}
\label{tab:efficiencies}
\begin{center}
\begin{tabular}{lcccc}
\hline
Channel & $\Upsilon$(1S) & $\Upsilon$(2S)  & $\Upsilon$(3S) & 4S \\
\hline
$\rho~\pi$                   &  7.8 & 6.6 & 6.0 & 5.6 \\
$K^*(892)~\bar{K}$           & 10.6 &10.0 & 9.6 & 9.5 \\
\hline
$\rho~a_2(1320)$             &  7.8 & 7.0 & 6.6 & 6.3 \\
$\omega~f_2(1270)$           &  7.7 & 6.9 & 6.5 & 6.1 \\
$\phi~f_2'(1535)$            & 10.1 & 10.1&10.2 &10.2 \\
$K^*(892)~\bar{K}_2^*(1430)$ &  5.3 & 5.2 & 5.1 & 5.1 \\
\hline
$b_1(1235)~\pi$              &  7.2 & 6.6 & 6.2 & 6.0 \\
$K_1(1270)~\bar{K}$          &  9.2 & 8.9 & 8.7 & 8.6 \\ 
$K_1(1400)~\bar{K}$          &  9.4 & 9.3 & 9.3 & 9.3 \\ 
\hline
\end{tabular}
\end{center}
\end{table}


\begin{table}[!htp]
\setlength{\tabcolsep}{0.4pc}
\catcode`?=\active \def?{\kern\digitwidth}
\caption{Number of events in each isospin-
and charge-conjugate-inclusive decay channel.}
\label{tab:signalyield}
\begin{center}
\begin{tabular}{lcccc}
\hline
Channel & $\Upsilon$(1S) & $\Upsilon$(2S)  & $\Upsilon$(3S) & 4S \\
\hline
$\rho~\pi$                   &  4 &  1 &  3 &  6  \\
$K^*(892)~\bar{K}$           & 18 &  2 &  4 & 15  \\
\hline
$\rho ~a_2(1320)$            & 29 &  8 & 10 & 47  \\
$\omega~f_2(1270)$           &  6 &  1 &  0 &  4  \\
$\phi~f_2'(1535)$            & 17 &  4 &  0 &  7  \\
$K^*(892)~\bar{K}_2^*(1430)$ & 16 &  6 &  5 & 23  \\
\hline
$b_1(1235)~\pi$              &  6 &  1 &  2 &  3  \\
$K_1(1270)~\bar{K}$          & 27 &  7 &  8 & 29  \\ 
$K_1(1400)~\bar{K}$          & 37 & 13 &  9 & 38  \\
\hline
\end{tabular}
\end{center}
\end{table}


\begin{table}[!htp]
\setlength{\tabcolsep}{0.4pc}
\catcode`?=\active \def?{\kern\digitwidth}
\caption{Range of scale factors for the 4S yield to 
the lower resonances for the combined
isospin-and charge-conjugate-inclusive decay channels,
including luminosity, efficiency energy dependence, and
range of $1/s^n$ scaling (see text).}
\label{tab:scalefactors}
\begin{center}
\begin{tabular}{lccc}
\hline
Channel & $\Upsilon$(1S) & $\Upsilon$(2S)  & $\Upsilon$(3S) \\
\hline
$\rho~\pi$                   & 0.23-0.45 & 0.12-0.17 & 0.15-0.17 \\
$K^*(892)~\bar{K}$           & 0.19-0.36 & 0.13-0.15 & 0.14-0.16 \\
\hline
$\rho ~a_2(1320)$            & 0.21-0.32 & 0.11-0.14 & 0.15-0.16 \\
$\omega~f_2(1270)$           & 0.21-0.32 & 0.12-0.14 & 0.15-0.16 \\
$\phi~f_2'(1535)$            & 0.17-0.26 & 0.12-0.13 & 0.14-0.15 \\
$K^*(892)~\bar{K}_2^*(1430)$ & 0.17-0.27 & 0.12-0.13 & 0.14-0.15 \\
\hline
$b_1(1235)~\pi$              & 0.20-0.31 & 0.11-0.14 & 0.15-0.16 \\
$K_1(1270)~\bar{K}$          & 0.17-0.27 & 0.12-0.13 & 0.14-0.15 \\
$K_1(1400)~\bar{K}$          & 0.16-0.25 & 0.12-0.13 & 0.14-0.15 \\
\hline
\end{tabular}
\end{center}
\end{table}


\begin{table}[!htp]
\setlength{\tabcolsep}{0.4pc}
\catcode`?=\active \def?{\kern\digitwidth}
\caption{Approximate background level estimate(s) from non-resonant 
$e^+e^+-$~annihilation scaled from the 4S data, and, where indicated
by a ``+'', cross-feed from other channels (see text), in the isospin- and
charge-conjugate-inclusive
decay channels.}
\label{tab:backgroundlevel}
\begin{center}
\begin{tabular}{lccc}
\hline
Channel & $\Upsilon$(1S) & $\Upsilon$(2S)  & $\Upsilon$(3S) \\
\hline
$\rho~\pi$                   & 3+2 &  1+0& 1+0 \\
$K^*(892)~\bar{K}$           & 5   &  2  & 2   \\
\hline
$\rho ~a_2(1320)$            & 15  &  7  & 8   \\
$\omega~f_2(1270)$           & 1   & 0.6 & 0.6 \\
$\phi~f_2'(1525)$            & 2   &  1  & 1   \\
$K^*(892)~\bar{K}_2^*(1430)$ & 6   &  3  & 4   \\
\hline
$b_1(1235)~\pi$              & 1   & 0.4 & 0.5 \\
$K_1(1270)~\bar{K}$          & 8+20& 4+8 & 4+4   \\
$K_1(1400)~\bar{K}$          & 10  & 5   & 6   \\
\hline
\end{tabular}
\end{center}
\end{table}


\begin{table}[!htp]
\setlength{\tabcolsep}{0.4pc}
\catcode`?=\active \def?{\kern\digitwidth}
\caption{Upper limits on number of signal events seen in each channel,
at 90\% confidence level, computed using the observed 4S yield
with the lower end of the range of scale factors ({\it i.e.} based on 
a $1/s$ cross-section dependence, resulting in a low background
prediction), and including
the statistical fluctuations of not only signal and background
but also the mean of the background.}
\label{tab:nrsigevtsupperlimits}
\begin{center}
\begin{tabular}{lccc}
\hline
Channel & $\Upsilon$(1S) & $\Upsilon$(2S)  & $\Upsilon$(3S) \\
\hline
$\rho~\pi$                   &  5.5 & 3.5 & 5.9 \\
$K^*(892)~\bar{K}$           & 22.1 & 4.0 & 6.1 \\
\hline
$\rho ~a_2(1320)$            & 27.6 & 8.1 & 8.8 \\
$\omega~f_2(1270)$           &  9.8 & 3.6 & 2.3 \\
$\phi~f_2'(1535)$            & 22.5 & 7.3 & 2.3 \\
$K^*(892)~\bar{K}_2^*(1430)$ & 18.7 & 8.0 & 6.4 \\
\hline
$b_1(1235)~\pi$              & 10.0 & 3.7 & 5.0 \\
$K_1(1270)~\bar{K}$          & 13.3 & 4.6 & 6.7 \\
$K_1(1400)~\bar{K}$          & 40   &14.5 & 9.2 \\
\hline
\end{tabular}
\end{center}
\end{table}


\begin{table}[thp]
\setlength{\tabcolsep}{0.4pc}
\catcode`?=\active \def?{\kern\digitwidth}
\caption{$\Upsilon$ branching fraction 68\% confidence intervals
and 90\% CL upper limits, 
in units of 10$^{-6}$,
for each isospin- and charge-conjugate-inclusive decay channel.
The first error is statistical and the second is systematic.
For the 90\%~CL upper limits, a low-end 4S scale factor was used,
whereas for the 68\%~confidence intervals the maximal scale
factor value was employed.      
Also shown is the statistical signal significance (in units of $\sigma$).
A ``$-$'' indicates less than 1$\sigma$ significance,
defined as
the number of Gaussian standard deviations above which
lies the probability that the background alone fluctuated up to the
observed number of events.
}
\label{tab:tableLast}
\begin{center}
\begin{tabular}{l|rr|rr|rr}
\hline
Channel & $\Upsilon$(1S) & & $\Upsilon$(2S) &  & $\Upsilon$(3S) & \\
        & Interval/Sig. & UL  & Interval/Sig. & UL  & Interval/Sig. & UL \\
\hline
\rule{0mm}{0.4cm}
$\rho~\pi$               & $-$ & 4 
                         & $-$ & 11 
                         & 9$_{-8}^{+7}\pm 1$~/~1.5 & 22 \\
$ K^*(892)~\bar{K}$      & 6$_{-2}^{+3}\pm 1$~/~3.6 & 11
                         & $-$ &  8
                         & $-$ & 14 \\
\hline
\rule{0mm}{0.4cm}
$\rho ~a_2(1320)$       & 9$\pm 4 \pm 1$~/~3.0 & 19
                        & $-$ & 24
                        & 8$_{-~6}^{+15}\pm 1$~/~1.3 & 30 \\
$\omega~f_2(1270)$      &  3$_{-1}^{+2}\pm 1$~/~2.6 & 7
                        & $-$ & 11
                        & $-$ &  8 \\
$\phi~f_2'(1525)$       &  7$_{-2}^{+3}\pm 1$~/~5.5 & 12
                        &  6$_{-3}^{+6}\pm 1$~/~3.0 & 17
                        & $-$ & 14 \\
$ K^*(892)~\bar{K}_2^*(1430)$ & 9$_{-4}^{+5}\pm 1$~/~3.0 & 19
                              & 11$\pm 8\pm 2$~/~1.6 & 32 
                              & $-$ & 28 \\
\hline
\rule{0mm}{0.4cm}
$ b_1(1235)~\pi$        &  3$\pm 2 \pm 1$~/~2.9 & 8
                        & $-$ & 12
                        & 5$_{-4}^{+9}\pm 1$~/~1.4 & 18 \\
$ K_1(1270)~\bar{K}$    & $-$ & 8
                        & $-$ & 11
                        & $-$ & 17 \\
                        
$ K_1(1400)~\bar{K}$    & 14$_{-3}^{+4}\pm 2$~/~5.6 & 23
                        & 16$_{-~7}^{+10}\pm 2$~/~2.9 & 33
                        &  7$_{-~5}^{+10}\pm 1$~/~1.5 & 22 \\
\hline
\end{tabular}
\end{center}
\end{table}



\begin{thebibliography}{99}  
  
\bibitem{pdg2002} Particle Data Group, K.~Hagiwara {\sl et al.},  
Phys. Rev. D {\bf 66} (2002) 010001.  
  
\bibitem{guandli} Y.F~Gu and X.H.~Li, Phys. Rev. D {\bf 63} (2001) 114019.  
  
\bibitem{BESPsiprimedecays} BES Collaboration, J.Z. Bai {\sl et al.},  
Phys. Rev. D {\bf 67} (2003) 052002.  
  
\bibitem{whatisHHC} S.J.~Brodsky and M.~Karliner,  
Phys. Rev. Lett. {\bf 78} (1997) 4682.  
  
\bibitem{BRODLEP} S.J.~Brodsky and G.P.~Lepage, Phys. Rev. D {\bf 24} (1981)  
2848.  
  
\bibitem{cleoiiidetector} CLEO Collaboration,
Y. Kubota {\sl et al.}, Nucl. Instrum. Meth. A {\bf 320} (1992) 66;
P.I. Hopman {\sl et al.}, Nucl. Instrum. Meth. A {\bf 384} (1996) 61;
I.~Shipsey {\sl et al.}, Nucl. Instrum. Meth. A {\bf 386} (1997) 37;  
P.~Skubic {\sl et al.}, Nucl. Instrum. Meth. A {\bf 418} (1998) 40;  
J.~Fast {\sl et al.}, Nucl. Instrum. Meth. A {\bf 435} (1999) 9;  
D. Peterson {\sl et al.}, Nucl. Instrum. Meth. A {\bf 478} (2002) 142.
  
\bibitem{RICHNIM} M. Artuso {\sl et al.}, 
Nucl. Instrum. Meth.A {\bf 461} (2001)  545.
 
\bibitem{GEANT} R.~Brun {\sl et al.}, CERN Report No. CERN-DD/EE/84-1,
1987; see also the web-site 
{\tt http://wwwasdoc.web.cern.ch/wwwasdoc/geant\_html3/geantall.html} .
  
\bibitem{FELDCOUS} G.J.~Feldman and R.D.~Cousins, Phys. Rev. D {\bf 57}  
(1998) 3873.  

\bibitem{BESpsi'} BES Collaboration, J.Z. Bai {\sl et al.},
Phys. Rev. Lett. {\bf 81}  
(1999) 1918.  

\end{thebibliography}
\end{document}